\documentclass{article}
\usepackage[export]{adjustbox}
\usepackage[utf8]{inputenc}
\usepackage{authblk}
\usepackage[margin=1.0in]{geometry}
\usepackage[square,sort,comma,numbers]{natbib}

\title{Parse Concurrent Structures: BST as an Example}
\author{Keren Zhou, Guocheng Niu, Wuzhao Zhang, Xueqi Li, Wenqin Liu}
\affil{Insititute of Computing Technology, Chinese Academy of Sciences}
\date{}

\usepackage{natbib}
\usepackage{graphicx}
\usepackage{graphics}
\usepackage{subcaption}
\usepackage{algorithm}
\usepackage{algorithmicx}
\usepackage{algpseudocode}
\usepackage{amsmath}
\usepackage{xcolor}
\usepackage{standalone}
\usepackage{tikz,pgf}
\usetikzlibrary{arrows,shapes,snakes,automata,backgrounds,petri}
\colorlet{kw}{yellow}
\def\infinity{\rotatebox{90}{8}}

\begin{document}
\tikzset{
  treenode/.style = {align=center, inner sep=0pt, text centered,
    font=\sffamily},
   error_n/.style = {red},
  circle_n/.style = {treenode, circle, black, draw=black,
    fill=white, text width=1.5em},
  circle_lock_n/.style = {treenode, circle, white,  draw=black,
    fill=white, text width=1.5em},
   lock_n/.style = {treenode, circle, white, draw,
   text width=1.5em}, 
  rec_n/.style = {treenode, rectangle, draw=black,fill=white,
    minimum width=1.5em, minimum height=1.5em},
  rec_lock_n/.style = {treenode, rectangle, draw=black,fill=white,
    minimum width=1.5em, minimum height=1.5em},
   comm_n/.style = {text centered},
	pil/.style={
           ->,
           thick,
           shorten <=2pt,
           shorten >=2pt,},
       pil2/.style={
           ->,
           dashed,
           shorten <=2pt,
           shorten >=2pt,}
}
\maketitle

\begin{abstract}
Designing concurrent data structures should follow some basic rules. By separating the algorithms into two phases, we present guidelines for scalable data structures, with a analysis model based on the Amadal's law. To the best of our knowledge, we are the first to formalize a practical model  for measuring concurrent structures' speedup. We also build some edge-cutting BSTs following our principles, testing them under different workloads. The result provides compelling evidence to back the our guidelines, and shows that our theory is useful for reasoning the varied speedup.
\end{abstract}

\section{Introduction}
As multi-core chips are widely used in commodity devices, designing concurrent structures has become a hot topic. These years, researchers focus on designing concurrent BSTs. Normally, sequential BSTs should be entirely locked when accessed by multi-threads. Concurrent BSTs leverage the property that modifications naturally happen in disparate places, therefore using finer-grained locks or flags could boost the parallelism.\\
To improve the performance of concurrent BSTs, there are several aspects of optimization. From the perspective of hardware interface, we could apply varied atomic operations, like compare-and-swap\cite{valois1995lock}, fetch-and-add\cite{gottlieb1981coordinating}. By the help of underlying system support, we can devise RCU\cite{mckenney2001read} and STM\cite{shavit1997software}. From the perspective of structures, external trees and internal trees are both available. To achieve greater disjoint-parallelism\cite{hoare1976parallel}, the locks which previous on the nodes could be moved to the edges.\\
ASCYLIB\cite{david2015asynchronized} is a concurrent structure library, including a bunch of different structures such as linked-list, hash-table, and BSTs. The core of ASCY is that concurrent structures should resemble their sequential counterparts. The author addresses that structures follow ASCY-compliant pattern use less power consumption, and achieve portable scalability that scale well under different workloads and platforms.\\
In this paper, we adopt the similar idea as ASCY to implement different concurrent BSTs. We compare their performance under various workloads and platforms, and propose our own principles of designing concurrent structures. Based on the Amdahl's law\cite{hill2008amdahl}, we present the first model to analyze speedup for concurrent structures. 

\section{Preliminary}
We view the BST as a dictionary to retrieve unique key-value pairs. There are three kinds of operations with the dictionary, we define them as follow:
\begin{itemize}
\item Search($key$). It calls $find$ operation to reach the corresponding leaf node, and returns $true$ if the $key$ matches, or it returns $false$.
\item Insert($key$). It begins to reach the candidate leaf node, if there's already such a $key$, it returns $false$. Otherwise it adds the $key$ into the dictionary.
\item Delete($key$). It begins to reach the candidate leaf node, if there's no such a $key$, it returns $false$. Otherwise it removes the $key$ from the dictionary.
\end{itemize}
To formalize the operations from the perspective of thread interactions, we propose a interface of our design in figure 1. Each operation will take ``snapshot'' of the tree by the $find$ routine, and adopt different consistency controller to manage contentions and start retry. The similar idea is used by \cite{herlihy2007simple}, to facilitate the performance under a little modification, the author use simple locks with some checks to develop a concurrent skiplist.\\
\begin{figure}[htbp]
\centering
\includegraphics[scale=0.6]{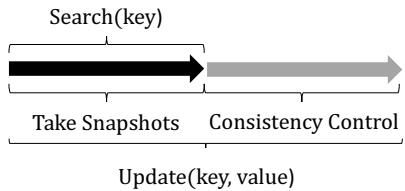}
\caption{Operation Interface}
\label{fig:my_label1}
\end{figure}
There are generally two types of structures in BSTs, one is the internal tree, the same as sequential BSTs; the other is the external tree, using more space but reduces contention to the leaf nodes. 
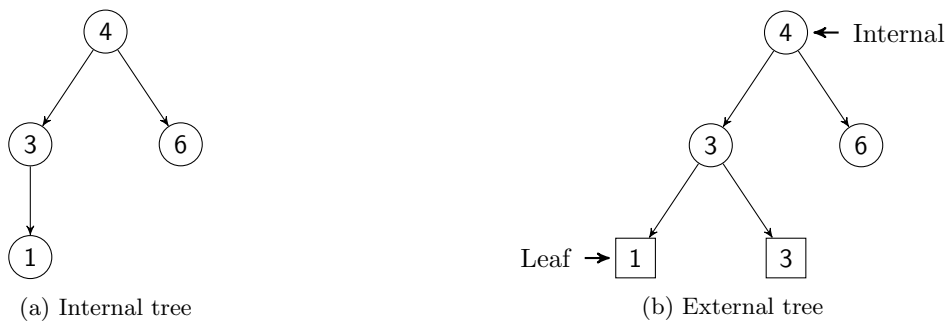
\begin{figure}[htbp]
 
\begin{subfigure}{0.5\textwidth}
\centering
\begin{tikzpicture}[->,>=stealth',level/.style={sibling distance = 2cm, level distance = 1.5cm}] 
\node [circle_n] (circle_4){4} 
            child{ node [circle_n] (circle_3){3} 
            	      child{ node [circle_n] (rec_1) {1} 
                }                         
    	     }
            child{ node [circle_n](circle_6) {6} }
; 
\end{tikzpicture}
\caption{Internal tree}
\label{fig:subim1}
\end{subfigure}
\begin{subfigure}{0.5\textwidth}
\centering
\begin{tikzpicture}[->,>=stealth',level/.style={sibling distance = 2cm, level distance = 1.5cm}] 
\node [circle_n] (circle_4){4} 
            child{ node [circle_n] (circle_3){3} 
            	      child{ node [rec_n] (rec_1) {1} 
                }
                    child{ node [rec_n](rec_3) {3}
                }                            
    	     }
            child{ node [circle_n](circle_6) {6} }
; 
\node[comm_n] (inter) at (1.5,0) {Internal}
edge[pil, bend right=0] (circle_4);
\node[comm_n] (leaf) at (-3.2,-3) {Leaf}
edge[pil, bend right=0] (rec_1);
\end{tikzpicture}
\caption{External tree}
\label{fig:subim2}
\end{subfigure}
 
\caption{Two types of BSTs}
\label{fig:image2}
\end{figure}
Figure 2 shows the different structures, where the external tree stores keys only in leaves, but the internal tree stores keys in both. The internal tree will be scaled up as the key buckets growing, whereas external tree renders better performance when the key range is small\cite{ramachandran2015castle}. We use the external tree to clearly our idea, and we believe that the same technique could be applying to the internal tree with little adapt. Our FEM-BST, using flag and mark indicators, is also very friendly to be adjusted into a lock-free version. \\
To avoid some special situations, we confine the key range  in $(-\infinity, \infinity)$ as describe in \cite{ellen2010non}. Figure 3 shows the initial structure, it is guaranteed that the initial three nodes will never be removed.
\begin{figure}[htbp]
\centering
\begin{tikzpicture}[->,>=stealth',level/.style={sibling distance = 2cm, level distance = 1.5cm}] 
\node [circle_n] (circle_4){$\infinity$} 
            child{ node [rec_n] (circle_3){$-\infinity$} }
            child{ node [rec_n](circle_6) {$\infinity$}}
; 

\end{tikzpicture}
\caption{Initial structure of our BST}
\label{fig:my_label5}
\end{figure}
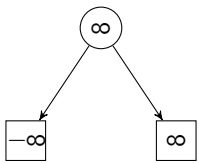
As illustrate in table 1, we implement 5 kinds of BSTs using different locks, all of which are CAS-based locks. Furthermore, we implement different consistency controllers to check whether the state is consistent during the modification. The ticket lock uses two numbers, ticket and version. The $version$ is to record the current version of the node, the $ticket$ is used for lock. If the current $ticket$ is not equal to the $version$, it indicates that the node is locked. The flag-marked lock simply use two boolean field. The flag field is to indicate whether the node is owned by a thread or not, the marked field is to denote whether the node is under the delete operation.
\begin{table}[htbp]
\centering
\begin{tabular}{c|c}
\hline
Name & Lock \\
\hline
BST & none \\
SYN-BST & synchronized \\
FN-BST & flag-based lock on node \\
FE-BST & flag-based lock on edge \\
FEM-BST & flag-mark-based lock on edge \\
TN-BST & ticket lock on node \\
\hline
\end{tabular}
\caption{The variant BSTs}
\label{tab:my_label1}
\end{table}
In this paper, we introduce the algorithm of FEM-BST. It has locks on nodes, but the performance is no different as the formal edge-based locks. We design fine checking mechanism to ensure the correctness and improve parallelism. 

\section{Algorithm}
\subsection{Search}
We use $ppred$ to denote the grandparent node, $pred$ to denote the parent node, and $curr$ for the current node. Moreover, $pright$ and $right$ represent the directions. As shown in the algorithm $FIND$, the operation goes from the root node to the corresponding leaf node, and returns a snapshot which includes 5 elements: $\{ppred, pright, pred, right, curr\}$ in figure 4. 
\begin{algorithm}
\caption{Find}
\begin{algorithmic}[1]
    \State $curr \gets root$
    \State $ppred, pred \gets null$
    \While {$curr \not= leaf$}
        \State $\text{taking snapshot}$
        \If {$curr.key < key$}
            \State $curr \gets curr.left$
        \Else
            \State $curr \gets curr.right$
        \EndIf
    \EndWhile
    \State \textbf{Return} $snapshot$
\end{algorithmic}
\end{algorithm}
The search algorithm is based on the optimistic strategy, which finds the node $was$ in the tree on the search path, other than the node $is$ currently in the tree. In fact, we can hardly implement such an algorithm that returns the result at the exact time-stamp. 
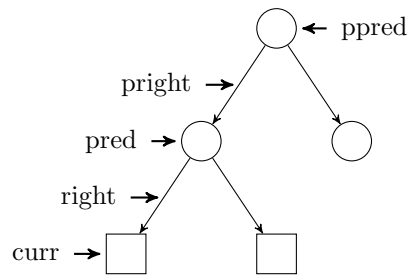
\begin{figure}[htbp]
\centering
\begin{tikzpicture}[->,>=stealth',level/.style={sibling distance = 2cm,
  level distance = 1.5cm}] 
\node [circle_n] (circle_1){} 
            child{ node [circle_n] (circle_2){} 
            	      child{ node [rec_n] (rec_1) {} 
                }
                    child{ node [rec_n](rec_2) {}
                }                            
    	     }
            child{ node [circle_n](circle_3) {} }
; 
\node[comm_n] (ppred) at (1.3,0) {ppred}
edge[pil, bend right=0] (circle_1);

\node[comm_n] (pright) at (-1.6,-0.75) {pright}
edge[pil, bend right=0]  (-0.5,-0.75);

\node[comm_n] (pred) at (-2.2,-1.5) {pred}
edge[pil, bend right=0]  (circle_2);

\node[comm_n] (right) at (-2.5,-2.25) {right}
edge[pil, bend right=0]  (-1.5,-2.25);

\node[comm_n] (curr) at (-3.2,-3) {curr}
edge[pil, bend right=0] (rec_1);

\end{tikzpicture}
\caption{Take snapshot from top to down}
\label{fig:my_label2}
\end{figure}

\subsection{Insert}
To begin with, the insert operation gets the snapshot in line 2. It compares whether the key is equal to the $curr$ node's. If there's such a key in the tree, it returns $false$. Otherwise it tries to lock the node, and handles some inconsistency situations. Figure 5 shows two inconsistency situations, one is that the $pred$ node is $marked$($line$ $9$), which indicates there's another operation deleting the $pred$ node; the other is when the $pred$ node is not linked to the $curr$ node($line$ $13$). Either of the situation indicates the operation has to retry to find the new corresponding node. There's no need to retry if the parent node is only locked, since it indicates that there's another node inserted in the other side while does not affect the current operation. The insert operation is guaranteed to be succeed when it locks the current node. Finally it constructs the node, and releases the lock.
\begin{algorithm}
\caption{Insert}
\begin{algorithmic}[1]
    \While {$TRUE$}
        \State $\{curr, pred, right\} \gets find(key)$
        \If {$curr.key == key$}
            \State \textbf{Return} \textit{FALSE}
        \EndIf
        \If {$!curr.tryLock()$}
            \State \textbf{Continue}
        \EndIf
        \If {$pred.marked$}\Comment{parent node already deleted}
            \State $curr.release()$
            \State \textbf{Continue}
        \EndIf
        \If {$right$ AND $pred.right \not= curr$ OR $!right$ AND $pred.left \not= curr$}
            \State $curr.release()$
            \State \textbf{Continue}
        \EndIf
        \State \text{Construct} $newParent$ \text{and} $insertNode$
        \If {$right$}\Comment{according to right flag}
            \State $pred.right \gets newParent$
        \Else
            \State $pred.left \gets newParent$
        \EndIf
        \State $curr.release()$
        \State \textbf{Return} \textit{TRUE}
    \EndWhile
\end{algorithmic}
\end{algorithm}

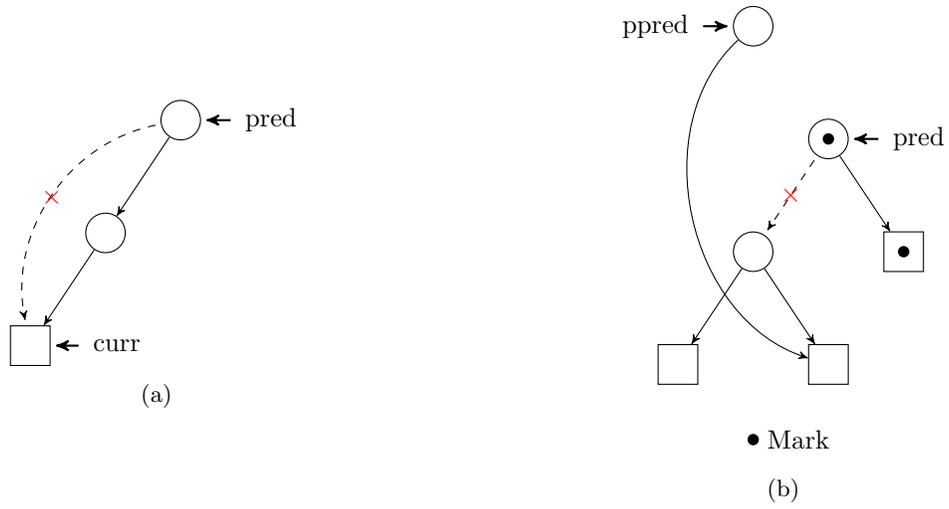
\begin{figure}[htbp]
 
\begin{subfigure}{0.5\textwidth}
\centering
\begin{tikzpicture}[->,>=stealth',level/.style={sibling distance = 2cm,
  level distance = 1.5cm}] 
\node [circle_n] (circle_1) at (0,0) {};
\node [circle_n] (circle_2) at (-1,-1.5) {};
\node [rec_n] (rec_1) at (-2,-3) {};

\node[comm_n] (curr) at (-0.85,-3) {curr}
edge[pil, bend right=0] (rec_1);

\node[comm_n] (curr) at (1.2, 0) {pred}
edge[pil, bend right=0] (circle_1);

\path[]
(circle_1) edge [] (circle_2)
		  edge [pil2,bend right = 45] node[error_n] {$\times$} (rec_1)
(circle_2) edge [] (rec_1);

\end{tikzpicture}
\caption{}
\label{fig:subim3}
\end{subfigure}
\begin{subfigure}{0.5\textwidth}
\centering

\begin{tikzpicture}[->,>=stealth',level/.style={sibling distance = 2cm,
  level distance = 1.5cm}] 
\node [circle_lock_n,tokens=1] (circle_1) at (0,0) {};
\node [circle_n] (circle_2) at (-1,-1.5) {};
\node [rec_lock_n,tokens=1] (rec_1) at (1,-1.5) {};
\node [rec_n] (rec_2) at (-2,-3) {};
\node [rec_n] (rec_3) at (0,-3) {};
\node [circle_n] (circle_3) at (-1,1.5) {};

\node [lock_n,tokens=1] (lock_1) at (-1,-4) [] {};
\node [comm_n] (lock_comm) at (-0.4,-4) [] {Mark};

\node[comm_n] (pred) at (1.2,0) {pred}
edge[pil, bend right=0] (circle_1);

\node[comm_n] (ppred) at (-2.3,1.5) {ppred}
edge[pil, bend right=0] (circle_3);

\path[]
(circle_1) edge [pil2,bend right = 0] node[error_n] {$\times$} (circle_2)
		  edge [] (rec_1)
(circle_2) edge [] (rec_2)
		  edge [] (rec_3)
(circle_3) edge [bend right = 60]  (rec_3)
;
\end{tikzpicture}
\caption{}
\label{fig:subim4}
\end{subfigure}
 
\caption{Two wrong situations of insert operation}
\label{fig:image3}
\end{figure}

\subsection{Delete}
Like the insert operation, the delete operation starts by getting the snapshot in Algorithm Delete $line$ $2$. However, it has to get an extra $ppred$ node and $pRight$ indicator, as it should move the grandparent's link to the sibling of the removed node. It first compares the key to the current node, if it is not equal, it returns $false$($line$ $4$). Otherwise it first tries to lock the $pred$ node($line$ $7$). It then checks the $ppred$ node's state($line$ $11$), ensuring the grandparent is neither marked or linked to another node. If it successes, it then begins to lock the current node($line 16$). Both the above locks has to be released once it detects inconsistency. After locking the parent node and the current node, it has to wait for the operation upon $sibling node$ to be finished($line $ $[34-48]$). The delete operation will successfully remove the node from the tree when it gets the correct sibling. 
\begin{algorithm}
\caption{Delete}
\begin{algorithmic}[1]
    \While {$TRUE$}
        \State $\{curr, pred, ppred, right, pright\} \gets find(key)$
        \If {$curr.key \not= key$}
            \State \textbf{Return} \textit{FALSE}
        \EndIf
        
        \State \text{Construct new $node$}
        \If {$pred.marked$ OR $!pred.tryLock()$} \Comment{check parent node}
            \State \textbf{Continue}
        \Else
            \State $pred.marked = \textit{TRUE}$
            \If {$ppred.marked$ OR $pRight$ AND $ppred.right \not\gets pred$ OR $!pRight$ AND $ppred.left \not= pred$}\Comment{check grandparent node}
                \State $pred.marked \gets \textit{FALSE}$
                \State $pred.relase()$
                \State \textbf{Continue}
            \EndIf
               
            \If {$!curr.tryLock()$} \Comment{check curr node}
                \State $pred.marked \gets \textit{FALSE}$
                \State $pred.release()$
                \State \textbf{Continue}
            \Else
                \State $curr.marked \gets \textit{TRUE}$
                \If {$right$ AND $pred.right \not= curr$ OR $!right$ AND $pred.left \not= curr$}
                    \State $curr.marked \gets false$
                    \State $curr.release()$
                    \State $pred.marked \gets false$
                    \State $pred.release()$
                    \State \textbf{Continue}
                \EndIf
                
                \If {$right$}\Comment{get sibling node}
                    \State $node \gets pred.left$
                \Else
                    \State $node \gets pred.right$
                \EndIf
                
                \While {$TRUE$}
                    \If {$right$}
                        \If {$node.lock$ OR $pred.left \not= node$}\Comment{The order cannot be changed}
                            \State $node \gets pred.left$
                            \State \textbf{Continue}
                        \EndIf
                        \State \textbf{Break}
                    \Else
                        \If {$node.lock$ OR $pred.left \not= node$}\Comment{The order cannot be changed}
                            \State $node \gets pred.left$
                            \State \textbf{Continue}
                        \EndIf
                        \State \textbf{Break}
                    \EndIf
                \EndWhile
            \EndIf
        \EndIf
        \algstore{myalg}
\end{algorithmic}
\end{algorithm}

\begin{algorithm}
\begin{algorithmic}[1]
\algrestore{myalg}
        \If {$pRight$}
            \State $ppred.right \gets node$
        \Else
            \State $ppred.left \gets node$
        \EndIf
        \State \textbf{Return} \textit{TRUE}
    \EndWhile
\end{algorithmic}
\end{algorithm}
Figure 6 shows the two situations the algorithm $Delete$ has to retry. Both happen on the grandparent node. For the situation b, we have to first check whether the node is released or not. Because once the sibling node is released, it could not be locked again when the parent node is locked. Therefore we do not have to retry from the root node. 
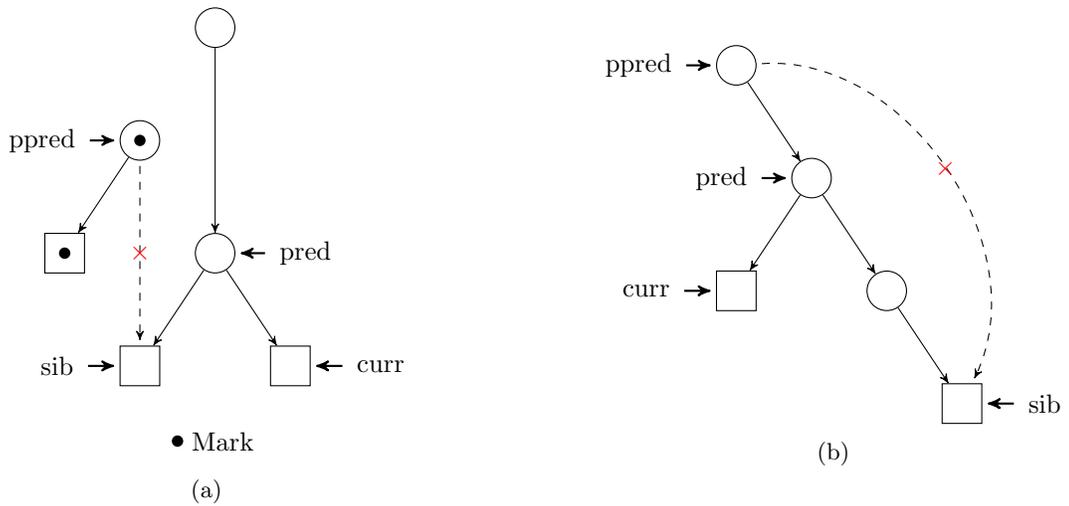
\begin{figure}[htbp]
 
\begin{subfigure}{0.5\textwidth}
\centering
\begin{tikzpicture}[->,>=stealth',level/.style={sibling distance = 2cm,
  level distance = 1.5cm}] 
\node [circle_n] (circle_1) at (0,0) {};
\node [rec_n] (rec_1) at (-1,-1.5) {};
\node [rec_n] (rec_2) at (1,-1.5) {};
\node [rec_lock_n,tokens=1] (rec_3) at (-2,0) {};

\node [circle_lock_n,tokens=1] (circle_2) at (-1,1.5) {};
\node [circle_n] (circle_3) at (0,3) {};

\node[comm_n] (pred) at (1.2,0) {pred}
edge[pil, bend right=0] (circle_1);

\node[comm_n] (ppred) at (-2.3,1.5) {ppred}
edge[pil, bend right=0] (circle_2);
\node[comm_n] (sib) at (-2.1,-1.5) {sib}
edge[pil, bend right=0] (rec_1);
\node[comm_n] (curr) at (2.2,-1.5) {curr}
edge[pil, bend right=0] (rec_2);

\node [lock_n,tokens=1] (lock_1) at (-0.5,-2.5) [] {};
\node [comm_n] (lock_comm) at (0.1,-2.5) [] {Mark};

\path[]
(circle_1) edge [] (rec_1)
		  edge [] (rec_2)
(circle_2) edge [pil2,bend right = 0] node[error_n] {$\times$}  (rec_1)
		  edge [] (rec_3)
(circle_3) edge[] (circle_1)
;
\end{tikzpicture}
\caption{}
\label{fig:subim5}
\end{subfigure}
\begin{subfigure}{0.5\textwidth}
\centering
\begin{tikzpicture}[->,>=stealth',level/.style={sibling distance = 2cm,
  level distance = 1.5cm}] 
\node [circle_n] (circle_1) at (0,0) {};
\node [rec_n] (rec_1) at (-1,-1.5) {};
\node [circle_n] (circle_2) at (1,-1.5) {};
\node [rec_n] (rec_2) at (2,-3) {};
\node [circle_n] (circle_3) at (-1,1.5) {};

\node[comm_n] (pred) at (-1.2,0) {pred}
edge[pil, bend right=0] (circle_1);

\node[comm_n] (ppred) at (-2.3,1.5) {ppred}
edge[pil, bend right=0] (circle_3);
\node[comm_n] (curr) at (-2.2,-1.5) {curr}
edge[pil, bend right=0] (rec_1);
\node[comm_n] (sib) at (3.1,-3) {sib}
edge[pil, bend right=0] (rec_2);

\path[]
(circle_1) edge [] (rec_1)
		  edge [] (circle_2)
(circle_2)  edge [] (rec_2)
(circle_3) edge[] (circle_1)
		edge [pil2,bend right = -60] node[error_n] {$\times$}  (rec_2)
;
\end{tikzpicture}
\caption{}
\label{fig:subim6}
\end{subfigure}
 
\caption{Two wrong situations of delete operation}
\label{fig:image4}
\end{figure}

\section{Correctness}
We first prove that our FEM-BST maintains the property during executions, then prove it is deadlock free, and point out the linearization points. The proof structure is similar as \cite{michael1996simple}.
\subsection{maintain structure}
We prove that the following invariants hold during the modification:
\begin{enumerate}
\item The root node is never removed.
\item The key field never changes.
\item The left child's key is always less than the parent node, the right child's key is always greater or equal to the parent node.
\item Once a parent node is locked and checked, the insert operation must succeed.
\item Once the parent node and current node are locked and checked, the delete operation must succeed.
\end{enumerate}
Proof:
\begin{enumerate}
\item The available key range is in $(-\infinity, \infinity)$, therefore the initial three nodes will never be retrieved by modifications.
\item An insert operation is finished by constructing two new nodes, linked with the existing node; A delete operation is finished by moving the grandparent pointer to the sibling of removed node. Hence the key field never changes.
\item Any modification upon the BST take a correct snapshot at a specific time-stamp, hence the $curr$ node must be a child of the $pred$ node. For a insert operation, the newly construct node follows the right direction; for a delete operation, the grandparent's pointer pointed to the existed child in the tree.
\item The insert operation first tries to lock the $curr$ node. After locking, no other operation could take upon the $curr$ node. It then checks whether the $pred$ node is marked. If it is marked, the $curr$ lock is released, the operation retries to find a new corresponding node. Otherwise the insert operation will succeed. 
\item The delete operation tries to lock the node in the following order: $pred->curr$. locks on the $pred$ node and $curr$ node ensure that other thread could see the marked status, thereby do not affect the current deletion. The $sibling$ node will finally be in a clean state, since there's finite set of modify operations. 
\end{enumerate}

\subsection{liveness}
We prove our FEM-BST is deadlock-free.\\
An important observation is that insert and delete operations locks the node from top to down order. The insert operation only locks the $curr$ node. The delete operation first locks the $pred$ node, and release it once it detects the child is locked. Therefore once contention is detected, the operations will be rolled back.\\
Another contention happens in the delete operation is when it tries to lock the sibling node.  An important observation is that insert and delete operations locks the node from top to down order. Because there's finite insert and delete operation, and any operation locks the node from top to down order, the $pred$ and $cur$ will not be violated. Finally the delete operatoin could get the sibling in isolation. Thus the FEM-BST is deadlock-free.
\subsection{linearization point}
\begin{itemize}
\item Insert. The linearization point of a successful insert operation is in Algorithm 2 $line$ $6$; the linearization point of a failure insertion operation is in Algorithm 2 $line$ $4$.
\item Delete. The linearization point of a successful delete operation is in Algorithm 3 $line$ $6$ and $line$ $10$; the linearization point of a failure delete operation is in Algorithm 3 $line$ $4$.
\end{itemize}

\section{Performance}
All of our BSTs in table 1 are implemented by Java, JDK 1.7. We set up the test by randomly inserting $\frac{bucket}{2}$ elements into the tree, and then running cases threads for 5s to ensure that elements are inserted and deleted multiple times. A similar idea is used in \cite{alistarh2014spraylist}. 
Experiments are performed on the platform of two Intel E5-2680 processors, 32 hardware threads with hyper-thread supported. The system is Red Hat Enterprise Linux Server release 6.3.\\
We compared different BSTs with respect to throughput, which is defined as the total number of operations completed per second. The number of threads was set from 1 to 32 and the bucket size is 10000 and 100000. We use two workloads: low-contention: 9$\%$ insert, 1$\%$ delete, 90$\%$search, and mid-contention: 20$\%$ insert, 10$\%$ delete, 70$\%$ search.  Figure 7 shows the result under different modification distribution.
\begin{figure}[htbp]
\includegraphics[width=1.0\textwidth, left]{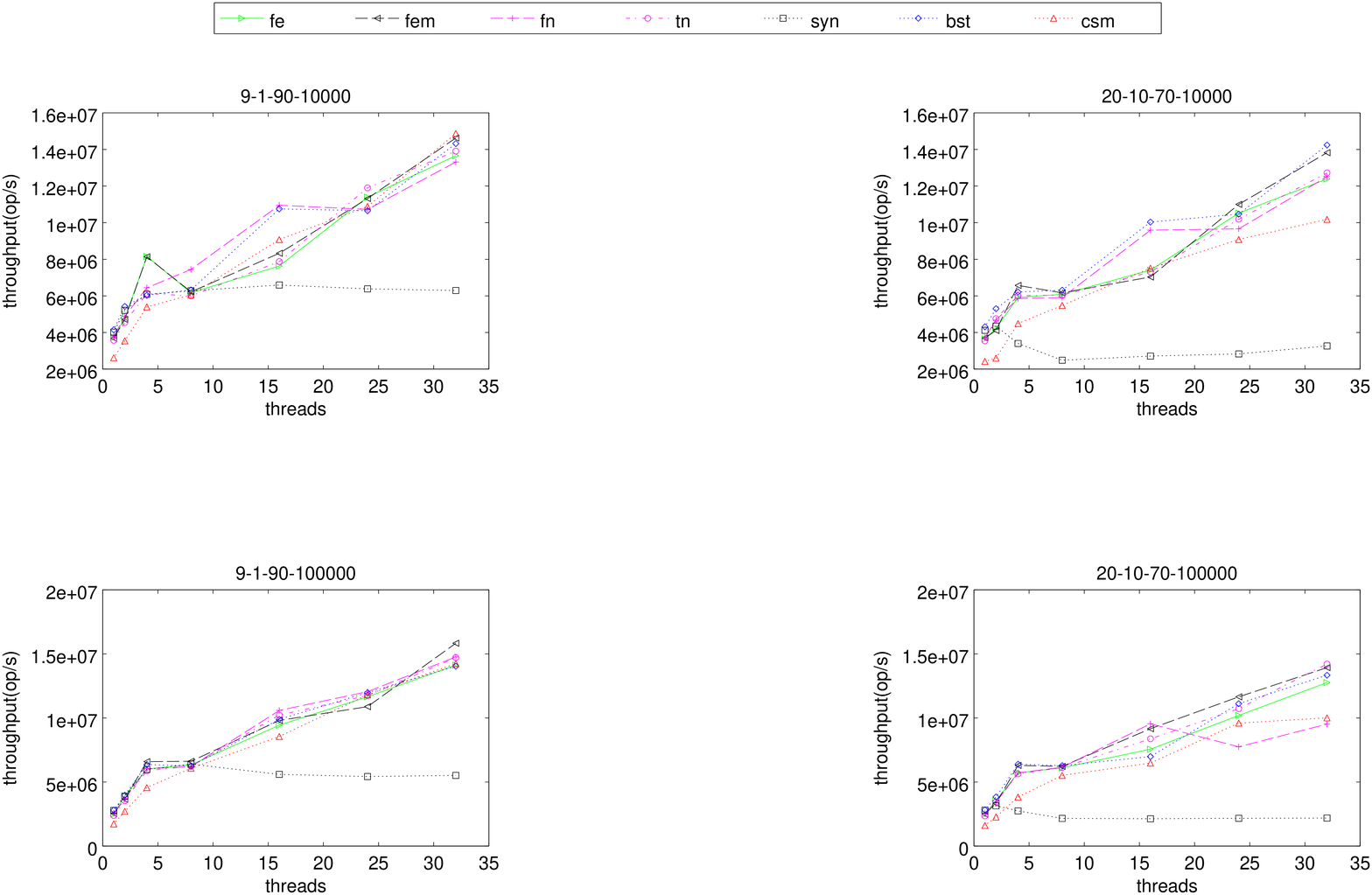}
\caption{Comparison of throughput of different BSTs under varied workload}
\label{fig:my_label}
\end{figure}
The result shows that our BSTs render similar performance as the unsynchronized-BST, which is used to stand for a possible upper-bound. The FN-BST is the least scalable one, since it need to lock more nodes than edge based BSTs. It is also worse than the TN-BST, which has to get little snapshot during searching. Our FEM-BST has the best performance, since it handles every possible inconsistency. We conclude our theory in the following part. 

\section{Principles}
We propose some concurrent data structure design principles according to our experiment and the interface in figure 1:
    \begin{enumerate}
    \item To achieve correctness, we must ensure either the snapshot or consistency controller exists.
    \item For the general lock-based algorithms, the more the amount of snapshot, the less complexity of consistency controller.
    \item For the lock-free algorithms, we might obtain both complex snapshot and consistency controller.
    \item Snapshot and consistency work together to affect single-thread performance. The higher the single-thread performance, the lower the progress conditions (parallelism).
    \end{enumerate}
Table 2 lists out our analyze of common concurrent techniques. $
Sycrhonized$ is for a coarse-grained version of sequential data structure, which handles the contention by locking the whole object. $STM$ utilizes parallelism by optimistic control strategy, and it should obtain a large amount of snapshot. $TicketLock$ is mentioned above, which implements the lock by version numbers. $FairLock$ represents fine-grained locks use a queued structure. $NonFairLock$ is implemented by flags. $Lockfree$ algorithms usually need very detailed design with thread interactions. $Waitfree$ algorithms are more strict than $Lockfree$ in progress condition, the only known waitfree structures are queue\cite{kogan2011wait} and linked-list\cite{timnat2012wait}.
        \begin{table}[htbp]
    \centering
    \begin{tabular}{c|c|c|c}
     \hline
     \textbf{Techniques} & \textbf{Parallelism} & \textbf{Snapshot} & \textbf{Consistency Controller}  \\
     \hline
     Sychronized & very low & very low & high \\
     STM & low & medium & medium\\
     TicketLock & medium & high & low\\
     FairLock & low & low & high(AQS framework)\\
     NonFairLock & medium & low & high\\
     Lockfree & medium & medium & high\\
     Waitfree & high & medium & very high\\
     \hline
    \end{tabular}
    \caption{Concurrent Techniques Comparison}
    \label{tab:my_label}
    \end{table}

\section{Model}
To the best of our knowledge, there's no any practical model fit for measuring the speedup of concurrent data structures. Here we present an analysis model to transform the initial amadal law for concurrent structures. 
\begin{equation*}
    speedup = \frac{1}{(1-p)+p/P}
\end{equation*}
The above equation is the most common known form of Amadal's law, where $p$ is the parallel ratio of a program. We assume $p = 1$ in concurrent structures, thereby the traditional model needs to be modified. We start from comparing the $workload$ of sequential part($w_{s}$) and parallel part($w_{p}$) of sequential structure to the concurrent structure where the parallel workload($w_{p}^{'}$) is different. 
\begin{equation*}
\begin{split}
speedup & = \frac{w_{s} + w_{p}}{w_{s} + w_{p} / P} \\
 & = \frac{w_{p}}{W_{p}(w_{snapshot}, w_{control}) / P(w_{snapshot}, w_{control})}
\end{split}
\end{equation*}
In the original equation, $P$ is defined as the number of processors, however in the concurrent structure, it is nearly impossible that all of the threads are taking into effect. Therefore we define $P(w_{snapshot}, w_{control})$ as a function which represents real parallelism. Furthermore, since new operations, snapshot and consistency control, are involved in the parallel version, we define $W_{p}(w_{snapshot}, w_{control})$ as the new parallel workload.
\begin{equation*}
\begin{split}
W_{p}(w_{snapshot}, w_{control}) &= w_{p} + w_{snapshot} + w_{control}\\
P(w_{snapshot}, w_{control}) &=P * (1 - c) * \alpha\\
0 &\leq c \leq 1
\end{split}
\end{equation*}
Where $c$ stands for the contention rate, $\alpha$ is the rate of taking effects on linearization points, $\beta$ is the rate of recording valid linearization points. Therefore $c$ is a experiment related variable, $\alpha$ and $\beta$ are algorithm related variables.
\begin{equation*}
\begin{split}
speedup &= \frac{P*(1-c)*\alpha}{1+\frac{w_{snapshot}}{w_{p}}+\frac{w_{control}}{w_{p}}}\\
\frac{1}{w_{snapshot}} &\leq \beta \leq 1\\
0 &\leq \alpha \leq \frac{w_{snapshot} * \beta}{w_{control}} \leq 1\\
\end{split}
\end{equation*}
The $\alpha$ factor is associated with the hardness of the sequential structure, where we define the hardness is proportional to the amount of adjust of the sequential part. Hence, the greater the a element has to communicate with others, the harder the structure, which means it needs more time to raise $\alpha$.
\begin{figure}
\centering
\includegraphics[width=0.7\textwidth]{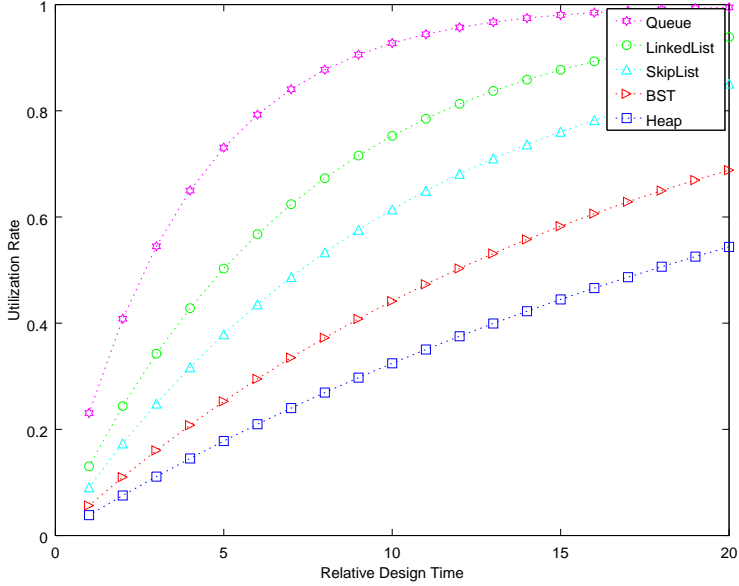}
\caption{
$
\alpha(t) = \frac{w_{snapshot} * \beta}{w_{control}}(-h^{-t}+1), 1 < h (\text{hardness of the structures})$
}
\label{fig:my_label8}
\end{figure}
Figure 8 demonstrates our measure of $\alpha$ factor. We have to pay much more amount of effort into ``hard'' structures. For instance, for the heap, we have to lock the whole path from root to leaf during modifications. Hence, to relax such a adjustment is difficult. However queue only need to modify the tail and head, therefore is easier to raise parallelism. 
\section{Conclusion}
We present a pattern of design concurrent data structures with a model to formalize the speedup measure. We also provide compelling evidence by measuring different kinds of BSTs under various workloads. An immediate discussion in the future would be implementing other structures such as skip-lists and heaps to illustrate our model. Another topic is to refine our model to measure the speedup accurately, and develop a software for practice. 

\bibliographystyle{plain}
\bibliography{references}
\end{document}